\documentstyle[emulateapj]{article}

\begin{document}

\def\EE#1{\times 10^{#1}}
\def\Msun{{~\rm M}_\odot}
\def\kms{\rm ~km~s^{-1}}
\def\lsim{\!\!\!\phantom{\le}\smash{\buildrel{}\over
  {\lower2.5dd\hbox{$\buildrel{\lower2dd\hbox{$\displaystyle<$}}\over
                               \sim$}}}\,\,}
\def\gsim{\!\!\!\phantom{\ge}\smash{\buildrel{}\over
  {\lower2.5dd\hbox{$\buildrel{\lower2dd\hbox{$\displaystyle>$}}\over
                               \sim$}}}\,\,}

\title{SN 1998bw @ late phases\footnote{Based on observations collected at the 
European Southern Observatory, La Silla and Paranal,
Chile.}}

\author{J.~Sollerman,\altaffilmark{2,3,4}
C.~Kozma,\altaffilmark{2}
C.~Fransson,\altaffilmark{2}
B.~Leibundgut,\altaffilmark{3}
P.~Lundqvist,\altaffilmark{2}
F.~Ryde,\altaffilmark{2}
P.~Woudt\altaffilmark{3}
}

\altaffiltext{2}{Stockholm Observatory, SE-133 36 Saltsj\"obaden, Sweden} 
\altaffiltext{3}{European Southern Observatory, Karl-Schwarzschild-Strasse 2, D-85784 Garching bei
M\"unchen, Germany}
 
\altaffiltext{4}{Send offprint requests to Jesper Sollerman; 
E-mail: jesper@astro.su.se}

\begin{abstract}
We present observations of the peculiar supernova SN~1998bw, which was
probably associated with GRB~980425. 
The photometric and spectroscopic evolution is monitored up to 500
days past explosion. We also present modeling based on spherically symmetric,
massive progenitor models and very energetic explosions. The models
allow line identification and clearly show the importance of
mixing. From the late light curves 
we estimate that $\sim 0.3-0.9\Msun$
of ejected $^{56}$Ni is required to power the supernova.

\end{abstract}


\section{Introduction}

Supernova (SN) 1998bw was born famous 
because of its positional and temporal coincidence with the gamma-ray
burst GRB~980425 (Galama et al. 1998a).
The unique nature of the associated radio source, in terms of the high 
luminosity immediately
after the explosion and 
the inferred relativistic expansion rate (Kulkarni et al. 1998), 
argues that SN~1998bw and GRB~980425 are related.
Independent of the connection to the GRB, SN~1998bw is an 
interesting supernova. It was optically luminous 
and displayed very high expansion velocities.
SN 1998bw was classified as a Type~Ic 
(Patat \& Piemonte 1998), 
and late time spectra reported here are
consistent with this classification.

Modeling of the early light curve and spectra suggested an extremely
energetic explosion 
of a massive carbon-oxygen star
(Iwamoto et al. 1998; Woosley, Eastman, \& Schmidt 1999). 
As the energy was
more than ten times 
that of a canonical
core-collapse supernova, the term 'hypernova' was suggested for this event
(Iwamoto et al. 1998). 
The mass of $^{56}$Ni needed to power the early light curve 
in these models, $0.5-0.7 \Msun$,
is much larger than the
$\lsim 0.1 \Msun$, typical for 'normal' core collapse SNe 
(Patat et al. 1994; Schmidt et al. 1994). 
This large mass of $^{56}$Ni 
for SN 1998bw was disputed by 
H\"oflich, Wheeler, \& Wang (1999), 
who found that the early light curve can
also be reproduced by an asymmetric 
SN~Ic ejecting only $0.2\Msun$ of $^{56}$Ni, 
given the right viewing angle and degree of asymmetry.

The models for SN~1998bw 
differ considerably
in terms of progenitor
mass, nickel mass and 
explosion energy. 
We therefore focus on the late time 
evolution, when the 
nucleosynthesis and density distribution can be studied directly. 
The total emission in the nebular phase is also less sensitive to
asymmetries, compared to the early light curve.

We present photometry and spectroscopy of SN 1998bw up to 500 days
past explosion.
We also compare these with results from spectral
synthesis based on various progenitor models. A more detailed
analysis will be given elsewhere.

\section{Observations, Reductions and Results}

\subsection{Photometry}
Photometry was obtained at eight epochs between 33 and 504 days past
explosion, 
using the ESO-3.6m telescope on La Silla and
the VLT/UT1 on Paranal. 
The images were bias subtracted and flat fielded using 
IRAF.

To obtain the supernova magnitudes we used 
DAOPHOT. 
The instrumental magnitudes were converted using
local standards from Galama et al. (1998b)\footnotemark.
At the earliest epochs, 
the errors are estimated as the standard deviations
in the magnitudes of the local standards. They also 
encapsulate
the neglect of color transformations.

\footnotetext{http://www.astro.uva.nl/$\sim$titus}

At later phases, the main uncertainty lies in the background
subtraction, as 
the supernova is superposed on an H~II region.
We constructed template PSFs using one or several stars, 
and the best PSF was chosen from
judgments of the subtracted images. 
This PSF was subtracted from the SN position with a range of
magnitudes, and
the supernova magnitude was
taken as the one which gave the smoothest background after subtraction.
Upper and lower limits were estimated 
by noting when the subtracted image showed a hole, or a
clear point source, at the supernova position.
The supernova magnitudes are given in Table~1 and plotted 
in Figure 1.

\subsection{Spectroscopy}

Spectroscopic observations were also obtained
to late phases (Fig.~2). At ESO-3.6m we used EFOSC2 with grisms 11 and
12, and at VLT/UT1 we used FORS1 with the 300-grisms.
All spectra were bias subtracted, flat fielded and 
wavelength calibrated. Flux calibration was done with a
spectrophotometric standard star, and the absolute fluxing was 
performed using the simultaneous $V$-band photometry. All spectra were taken
close to 
the parallactic angle to reduce differential refraction.

The spectral evolution of SN~1998bw from day 33 to day 504 is shown in
Figure 2. 
The evolution of the spectra in the nebular phase is fairly slow.
The [O~I] $\lambda\lambda 6300,6364$ are the strongest lines, and
the 7300 and 8600~\AA~features, likely to be due to
Ca~II, decrease in strength relative to the [O I] lines. 
Other prominent features are seen at $\sim4570$ and 5890 \AA, and are
probably due to Mg I] (or [Fe III]) and Na I D (and possibly He I), 
respectively.

\section{Discussion}
\subsection{General considerations}

A lower limit of $0.22\Msun$ of $^{56}$Ni for SN 1998bw 
was achieved by McKenzie \& Schaefer (1999) by comparing the 
$V$-magnitude at 170 days to that of SN 1987A.
While the light curve of 
SN 1987A closely followed the decay rate of $^{56}$Co, suggesting full 
trapping of the $\gamma$-rays, the light curve of SN 1998bw
fell substantially faster. 
The assumption of full trapping at 170 days in SN 1998bw 
used by McKenzie \& Schaefer is clearly too conservative.
In fact, 
the light curve of SN 1998bw is similar to that of many other SNe Ib/c 
(Clocchiatti \& Wheeler 1997; Sollerman, Leibundgut, \& Spyromilio 1998), 
and can be reproduced by a simple model of a radioactively powered ejecta 
leaking
$\gamma$-rays due to homologous expansion (Fig.~1). 
If the decay would follow full trapping 
from day 64 and onward, 
the flux in the $V$-band at day 170 would be 
2.4 times brighter than actually observed (Fig.~1). 
Assuming full trapping at day 64, 
at the beginning of the fast decay, then provides an upper limit of 
$\sim41\%$
on the trapping at day 170,
increasing the lower limit of the Ni-mass 
to $\sim0.5\Msun$.

The above estimate depends, however, on the 
bolometric correction and its evolution for these supernovae.
That the bolometric decay rate is 
similar to that of the $V$-band at this epoch is indicated by
the simple estimates of the bolometric luminosity 
by McKenzie \& Schaefer (1999). 
This is also confirmed by our 
detailed modeling (see below) and supported by the
fit to the $V$-curve by the simple $\gamma$-leakage model. 
However, the bolometric correction could still be different for SNe 
1987A and 1998bw. 
If SN 1998bw emitted a smaller fraction of its bolometric luminosity in the 
optical range than SN 1987A, its Ni-mass would be overestimated. 

Adopting a distance of 35 Mpc ($H_{0}=72)$, 
a reddening of $E(B-V)=0.06$ (Schlegel, Finkbeiner, \& Davis 1998), and
integrating our spectrum at day 141, gives a luminosity 
($L_{\rm opt}=3.5\times10^{41}$ ergs s$^{-1}$)
that requires 
$0.1\Msun$ of $^{56}$Ni. This limit clearly has to be corrected upwards as it 
assumes full $\gamma$-ray trapping, and that our spectrum contains all the
flux emitted by the supernova.
With the same argument as above, no more than $48\%$ 
of the energy is deposited at 141 days. 
Furthermore, 
detailed models (see below) show that the observed spectrum contains
only $71\%$ of the total flux. 
These corrections 
indicate that the lower limit must be increased to $0.3\Msun$.
The dominant errors are the uncertainties in bolometric correction 
($\sim10\%$),
distance, and reddening for SN 1998bw. Allowing for a 15$\%$ 
uncertainty in the distance scale, and $\pm0.1$ mag in $A_{V}$, make the
absolute fluxes (and the Ni-mass) uncertain by $\sim35\%$.

\subsection{Detailed modeling}

To obtain more quantitative estimates of the Ni-mass, detailed modeling is 
required. 
Our analysis uses an updated version of the 
spectral code in Kozma \& Fransson (1998a,~1998b). 
At the epochs discussed here, the ejecta emission is powered by 
the radioactive decay of $^{56}$Co.
The deposition of the resulting 
$\gamma$-rays and positrons
is modeled in detail according to
Kozma \& Fransson (1992). The ionization and heating is 
balanced by thermal and radiative processes, producing the 
observed optical and infrared emission.
This time-dependent code 
gives good agreement with light curves and late time
spectra for SN~1987A (Kozma \& Fransson 1998a, 1998b; Kozma 2000).

As input models we use different combinations of unmixed and
artificially mixed models of SN~1998bw, based on the CO138-model
calculated by Iwamoto et
al. (1998) and the CO6-model from Woosley et al. (1999).
Both models are spherically symmetric explosions of
massive carbon-oxygen stars.
The CO138-model has a mass $M_{\rm CO}$ = 13.8$\Msun$, 
explosion energy $E_{\rm exp}=3\EE{52}$ ergs, and mass of $^{56}$Ni, 
$M$($^{56}$Ni) = 0.7$\Msun$. 
Such a carbon-oxygen core corresponds to
a zero age main sequence (ZAMS) star of $M_{\rm ZAMS} \sim 40\Msun$.
The CO6C-model with $M_{\rm CO}$ = 6.55$\Msun$,
$E_{\rm exp} = 2.2\EE{52}$ ergs, and $M$($^{56}$Ni) = 0.47$\Msun$ 
corresponds to a $M_{\rm ZAMS} \sim 25\Msun$ star.

In Figure 3 we show our modeled spectra
together with the observations at 141 days. 
From the observed line profiles we  
conclude that mixing is important. 
Calculations based on non-mixed models give
very broad and flat-topped line profiles 
which are not seen in the observations.
In the mixed models, 
we have also decreased the expansion velocities to get a better
agreement with the observed line widths.
A mixed model with the original velocities results in nicely peaked,
but too broad, line profiles. 
A decrease in velocity results in 
an increase of the densities given by the original explosion models.
We have increased the densities by factors of 10 -- 100, 
except in the Fe-rich regions where it is kept unchanged (see below).
These changes also affect the kinetic energy in the 
modified models. In the layers below $ \sim 12\,000 \kms$, which
is the only material which contributes significantly to the 
emission at late times, the kinetic energies are 
$2\EE{51}$ ergs (CO6) and $5\EE{51}$ ergs (CO138), respectively.
Note that our analysis is not sensitive to the 
high velocity gas, which might also contain a significant amount of energy. 
Early time analysis is therefore more reliable for determining the 
total kinetic energy of the explosion.
As seen in Figure 3, the synthetic line profiles are 
still somewhat broader than the observed.
We can overcome this by decreasing the velocities even more, but
the high density
in the oxygen-rich regions then results in strong O~I lines at
$\lambda\lambda~5577,7774,9265$, which are not seen in the observations. 

Most features in the observed spectrum have a counterpart in the modeled
spectra, and in Figure 3 we include some of the line identifications. 
The $4000-5500$~\AA~regime is problematic, and
preliminary results from improved models suggest
that most of these 
features are a mixture of Fe~II and Fe~III lines.
The models presented here have too low a degree of ionization, 
with mostly Fe~I and Fe~II emission. A higher density in the Fe-emitting 
material would result in even lower ionization, which is 
why we do not increase the density in the Fe-rich regions. 
The ionization can instead be increased 
by decreasing the density in these regions.
Scattering and fluorescence may also play a significant role
in understanding these features. 
We therefore regard the Fe~I emission as dubious.
\clearpage

\subsection{The Light Curve}

In Figure 1 we show the light curves from our calculations of 
the mixed CO6- and CO138-models (solid lines), 
and from the original, non-mixed models (dotted lines).
The CO138-model is brighter than CO6.
This is not only due to the higher Ni-mass,
but also a result of the higher ejecta-mass ($M_{\rm ej}$), and lower 
expansion velocity
($v_{\rm max}$), 
giving a larger optical depth to the $\gamma$-rays ($\tau_\gamma$); 
$\tau_\gamma \propto M_{\rm ej} / v_{\rm max}^2$. 

The effect of changing the expansion velocities can also be seen 
in Figure 1. While the two CO6-models 
have the same Ni-mass and ejecta-mass, they differ in
expansion velocity. The mixed models 
are somewhat brighter than the non-mixed models, due to the higher
$\gamma$-ray optical depth 
for the models with lower velocities. 
In the original CO6-model 
we find that at 200~(400,~600) days only $\sim 7.2~(4.4,~3.9) \%$ 
of the energy released in the $^{56}$Co-decays is deposited in the ejecta.
Out of this energy $\sim 47~(78,~87) \%$ is from the kinetic energy of
the positrons. 
For the mixed CO6-model with decreased velocities, 
$\sim 11~(5.7,~4.5) \%$ of the energy is deposited.
 
From the light curves,
and within the framework of our modeling,
we conclude that the nickel mass needed to power the late emission 
is indeed very high.
For the two models that best reproduce the late time
observations, we calculate the deposition of the radioactive energy, and thus 
the required mass of nickel. The CO6-model requires $0.9\Msun$ of
$^{56}$Ni to maintain the luminosity at $340-410$ days. 
The CO138-models, which traps 
$10\%$ of the energy at 400 days, requires $0.5\Msun$.
The errors on these estimates are dominated by the uncertainty in distance and 
reddening stated in \S 3.1.

H\"oflich et al. (1999) found that the early light curve of
SN 1998bw could be explained in terms of a more 'normal' SN Ic, which ejected
0.2$\Msun$ of $^{56}$Ni, given the right degree of asymmetry.
However, at the nebular phase studied here the ejecta are optically thin, and
for a given density structure the luminosity is not as sensitive to 
asymmetries as is the early light curve. 
By modeling the late time line 
profiles, providing a measure of the energy input with velocity, we 
get a handle on the distribution of the ejecta, and can therefore model the 
gamma-deposition self-consistently.

The Ni-mass determined for the CO138-model is close to the lower limit from
\S 3.1. This is because this massive model closely represents the maximum 
trapping allowed by the observed light curve.
Also, as the mixed CO6-model is already dominated by positrons at 400 days, 
the estimated Ni-mass for this model must be close to the upper limit.

While SN 1998bw ejected a large mass of $^{56}$Ni,
a few other supernovae have shown very low
amounts of ejected $^{56}$Ni 
(SN 1994W [$\lesssim 0.015\Msun$, Sollerman, Cumming, \& Lundqvist 1998]
and SN 1997D [$\sim 0.002\Msun$, Turatto et al. 1998]). The
iron yield from core-collapse supernovae thus seems to vary substantially.

At very late phases ($>400$ days), the low energy input 
means low temperatures for the ejecta, and most of the emission in our models 
comes out in the infrared. The optical light curves no
longer directly follow the bolometric light curve.
After 400 days the model light curves
drop faster than the observations. A reduced density in the Fe-rich material 
(\S 3.2) can increase the temperature and boost in particular the $V$-curve 
at late phases.
 
Another possible source for the powering at late phases is
interaction with a circumstellar medium (CSM).
The absence of H and He in the spectra means that 
substantial amounts of such gas should be present. 
Models of the radio emission
indicate that some interaction is already taking place 
(Li \& Chevalier 1999),
although the CSM density in these models is probably too low to 
affect the optical light curve, as the wind velocity is high 
($\sim 2000\kms$). 
We see no significant spectroscopic signature 
that indicates CSM interaction, in our late spectra.

\section{Summary}
We present photometry and spectroscopy of SN 1998bw up to 500 days 
past explosion. Starting with 
two different spherically symmetric explosion models, 
we calculate late time spectra and light curves using 
a detailed spectral synthesis code to model our observations.
We conclude that mixing is important, and we
also have to alter the density distribution macroscopically 
to obtain reasonable agreement with the line profiles in the 
observed spectra.

Representing the supernova density profile with the distribution which best
reproduces the late time observations, we calculate the 
deposition of the radioactive energy, and determine how much $^{56}$Ni 
is required to produce the late time luminosity.
We find that a large amount, roughly half a solar mass, is needed to power the 
light curve one year after the explosion, although the
exact number depends on the chosen explosion model.

Evidence for asymmetries in SN 1998bw comes from early polarization 
measurements (Kay et al. 1998). 
The modeling presented here is restricted to the available symmetric models,
although the importance of mixing 
may point to an asymmetric explosion. 
Whether asymmetric models can reproduce the
observations is of interest for future studies.

\acknowledgements
We thank the Swedish Natural Science Research Council,
the Swedish National Space Board and the Knut and Alice Wallenberg
Foundation for support.
J.~S. is grateful to the grant from ESO Director's
Discretionary fund, and from Holmbergs and Hiertas funds.

\begin{deluxetable}{lccccc}

\tablewidth{0pt}
\tablecaption{supernova magnitudes}
\tablehead{Epoch\tablenotemark{a} & Telescope\tablenotemark &
$V$ & $R$ & $I$}
\startdata

33& 3.6m  &  14.69$\pm$0.05 & 14.31$\pm$0.05  & 14.26$\pm$0.05 \\

141& 3.6m     &  17.34$\pm$0.05        & 17.06$\pm$0.05 & 16.78$\pm$0.05 \\

215& 3.6m   &  18.63$\pm$0.05        & 18.10$\pm$0.05  & 17.92$\pm$0.05 \\

340& 3.6m   &  20.7$^{+0.3}_{-0.2}$  & 19.9$\pm$0.15  & 19.65$\pm$0.2 \\

357& VLT & 21.0$^{+0.25}_{-0.15}$& 20.2$^{+0.15}_{-0.10}$& 20.0$\pm$0.15 \\


414& VLT  & 21.7$\pm$0.3 & 21.0$\pm$0.20 & 20.9$^{+0.3}_{-0.2}$ \\

471& 3.6m  &21.9$^{+0.5}_{-0.2}$  & 21.6$\pm$0.30& 21.5$^{+0.3}_{-0.2}$ \\

504& VLT & 22.3$^{+0.4}_{-0.2}$ & 22.1$\pm$0.30& 21.8$^{+0.5}_{-0.2}$\\

\enddata

\tablenotetext{a}{Epoch in days past April 25, 1998, which we take as
the date of explosion.}

\end{deluxetable}

\clearpage
\begin{figure*} \centering \vspace{16.0 cm}
\includegraphics{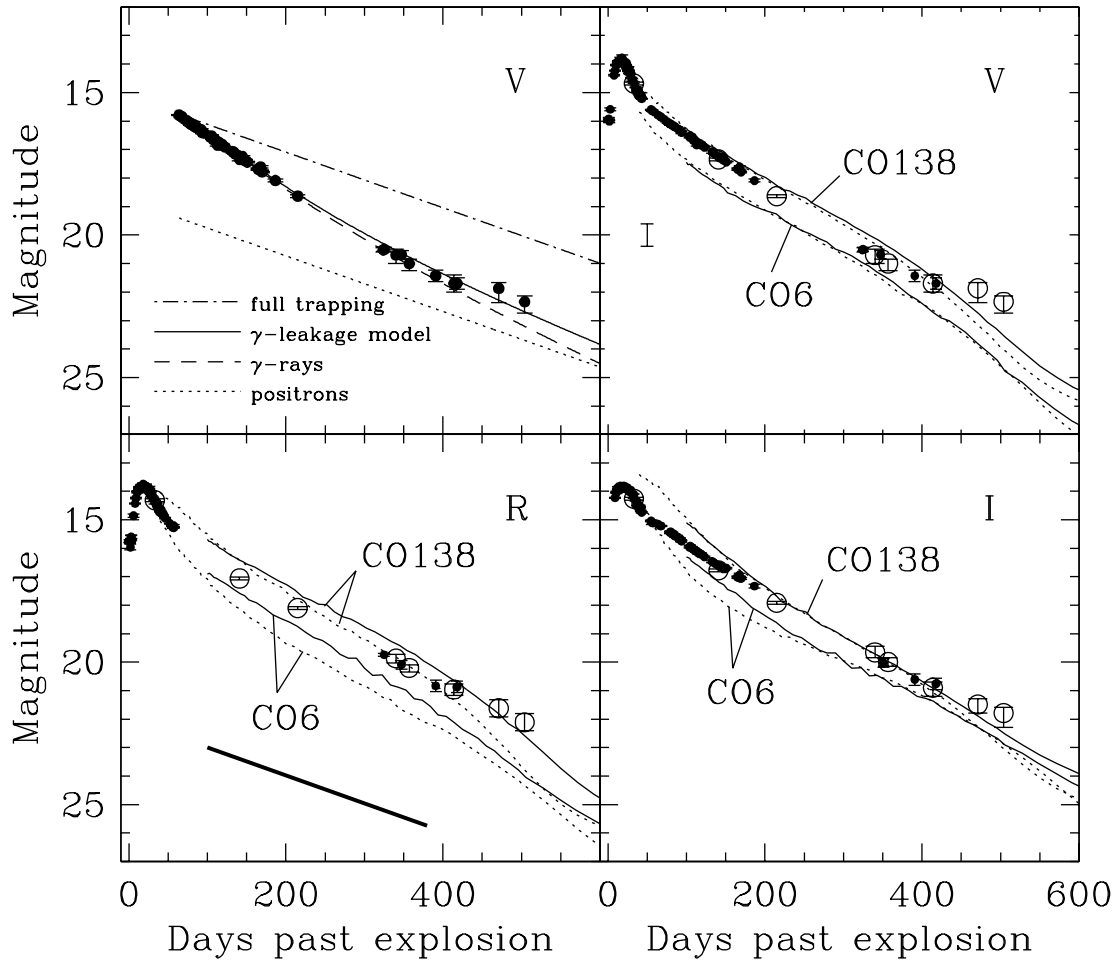}
\caption{
The upper left panel shows a simple 
model (solid line) 
compared to the $V$-band data 
described below. The flux is given by 
$e^{-t/111.3}\times(1-0.965~ ~e^{-\tau})$, where
111.3 days is the decay time of $^{56}$Co, and the $\gamma$-ray optical depth
$\tau = (t_{1}/t)^2$ (here $t_{1} = 120$ days). The dashed curve is the 
$\gamma$-ray contribution and the dotted line the positron contribution. 
The dashed-dotted
line shows the expected flux for full trapping from day 
64 and onward.
The other panels show the $V$, $R$, and $I$ light curves, respectively,
including, in addition to our data (large empty circles), 
observations from Galama et al. (1998a)
(1-- 57 days past explosion), 
McKenzie \& Schaefer (1999) (days 63 -- 187), and Patat et al. (2000) 
(days 324 -- 417).
The models included are CO138 and CO6 with mixed ejecta and decreased
velocities (solid lines), 
and the original CO138- and CO6-models (dotted lines).
The $\Delta m=\pm0.39$ error bar in the $V$-panel shows the uncertainty
in distance and reddening. The thick solid line in the $R$-panel 
shows the decay rate of $^{56}$Co for full $\gamma$-ray trapping.
}
\label{lightc}
\end{figure*}

\newpage
\clearpage
\begin{figure*} \centering \vspace{14.0 cm}
\includegraphics{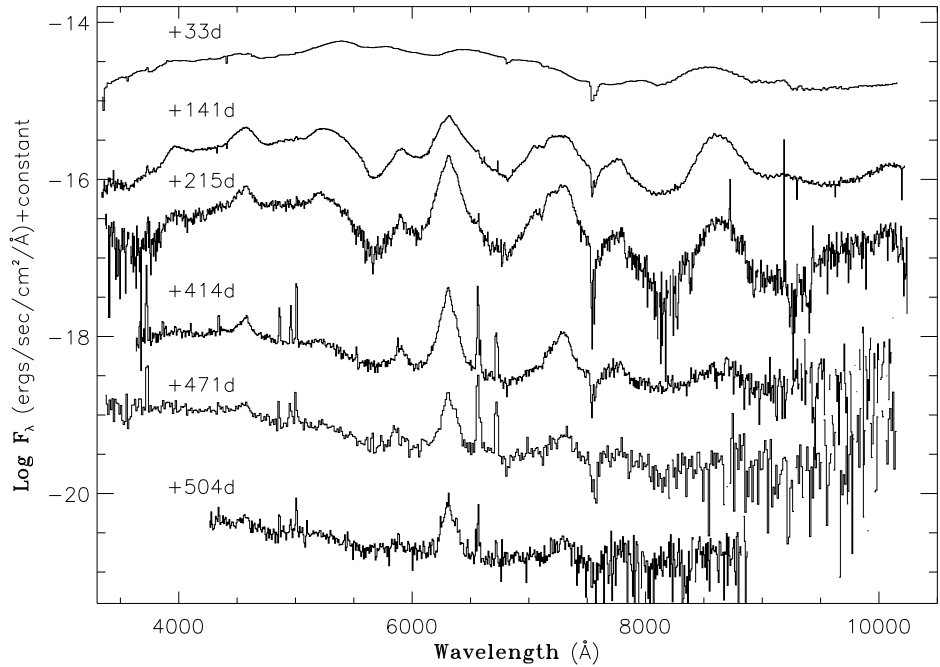}
\caption{
The spectral evolution of SN 1998bw from day 33 to day 504.
For clarity the spectra have been shifted by a constant factor, 
$\Delta$ log $F$$_{\lambda}$ :
+33$^{d}$ (0.0), 141$^{d}$ (--0.15), 215$^{d}$ (--0.5),
414$^{d}$ (--1.0), 471$^{d}$ (--2.0), and 504$^{d}$ (--3.2).
The wavelength scale has been corrected for the velocity of the parent
galaxy, 2550~$\kms$.
}
\label{specobs}
\end{figure*}

\newpage
\begin{figure*} \centering \vspace{14.0 cm}
\includegraphics{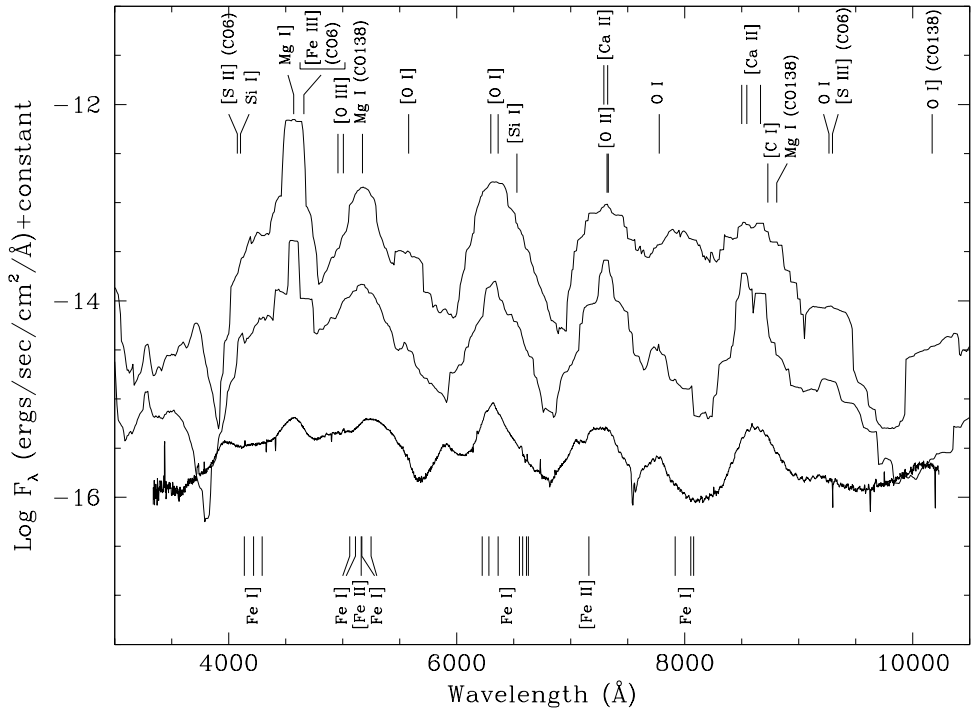}
\caption{
Synthetic spectra based on the  CO138-model (upper) and the 
CO6-model (middle) together with the observed spectrum (lower) at 141 days.
The CO138- and CO6-models were artificially mixed with decreased
velocities as described in the text. 
For some of the line identifications we 
indicate in which of the two models the line is important. 
The Fe~I and Fe~II lines are strongest in the CO138-model. 
The synthetic spectra have been shifted vertically by  
$\Delta$~log~$F$$_{\lambda}$ = --1.5 (CO6) and 
$\Delta$~log~$F$$_{\lambda}$ = --2.0 (CO138).
}
\label{spectra_mod}
\end{figure*} 

\end{document}